# EEG Wheelchair for People of Determination

[1]Mariam AlAbboudi, [1]Maitha Majed, [1]Fatima Hassan, [2]Ali Bou Nassif

[1]Department of Electrical and Electronics Engineering, University of Sharjah
[2]Department of Computer Engineering, University of Sharjah

Sharjah, United Arab Emirates

*Abstract*—the aim of this paper is to design and construct an electroencephalograph (EEG) based brain-controlled wheelchair to provide a communication bridge from the nervous system to the external technical device for people of determination or individuals suffering from partial or complete paralysis. EEG is a technique that reads the activity of the brain by capturing brain signals non- invasively using a special EEG headset. The signals acquired go through pre-processing, feature extraction and classification. This technique allows human thoughts alone to be converted to control the wheelchair. The commands used are moving to the right, left, forward, and backward and stop. The brain signals are acquired using the Emotiv Epoc headset. Discrete Wavelet Transform is used for feature extraction and Support Vector Machine (SVM) is used for classification.

## I. INTRODUCTION

The aim of this paper is to design a prototype to help those who cannot use their muscles to move by creating a prototype that works by brain signals. We focused on the simplicity of the project design where it will only require wearing the headset and think of the direction desired. The system is based upon a technique called Brain-Computer Interface BCI. It is a technique that provides a connection between the brain and the controller to translate the individual brain commands into actions. The system uses the brain electrical signals which are EEG signals where these signals can be used in many applications such as controlling a wheelchair, robot…etc. Electroencephalography (EEG) is the study of the recording of brain electrical signals that are used in analyzing neurocognitive processing of human activity to allow the user to have an insight of the real processing of the brain. Brain signals are collected using electrodes that are placed on the scalp with a conductive paste to allow the collecting of brain signals. The EEG frequency ranges are divided into five different frequency bands such as Delta (up to 4 Hz), Theta (4-8Hz), Alpha (8-15 Hz), Beta (15-32 Hz) and Gamma (greater than 32 Hz) [1].

The remaining of the paper is structured as follows: Section II presents related work where Section III proposes the system design. Sections IV and V demonstrate software and hardware design, respectively. Section VI presents the results and finally, VII concludes the paper and suggests future work.

## II. RELATED WORK

Brain-controlled wheelchairs have been implemented by many researchers and engineers in the field. "A Brain-Controlled Wheelchair to Navigate in Familiar Environments," was implemented in 2010 by Brice Rebsamen and his colleagues at the University of Singapore and Imperial College of London which is based on P300 interface and they got an average error rate of 12% [2]. In addition, Shedeed et al. developed "Brain EEG signal processing for controlling a robotic arm," in Ain Shams and Benha University in Egypt in 2008 where they achieved an error rate of 9% for 3 actions [3]. Zaki et al. also implemented a "Home Automation using EMOTIV: Controlling TV by Brainwaves," in 2018 at King Fahad University of Petroleum and Minerals in Dhahran, Saudi Arabia [4]. Additionally, Hamzah et al. also developed "Special Issue EEG signal classification to detect left and right command," a system based on Artificial Neural Network (ANN) algorithm in 2017 [1].

The main contribution of our work is that we fully implemented a system composed of software and hardware that works based on EEG signals. Our system works on 5 commands; right direction, left direction, forward, backword and stop with high accuracy.

## III. PROPOSED SYSTEM DESIGN

The aim of this project is to design a brain-controlled wheelchair for individuals suffering from partial or complete paralysis by constructing a wheelchair prototype to demonstrate the project concept which has five classified commands which are moving to the right, left, forward, backward or stop. The data is being collected using an EEG Emotiv Epoc headset to be able to control the wheelchair, the signals acquired must go through pre-processing, feature extraction and classification. Finally, the classified command will be sent to the controller to execute the command.

### A. Data acquisition

Raw data from the brain are acquired using an EEG headset called Emotiv EPOC+ headset according to the 10/20 international system for electrodes positioning as shown in the Figure 1. The Emotiv headset has 14 channel electrodes including two reference electrodes which are AF3, F7, F3, FC5, T7, P7, O1, O2, P8, T8, FC6, F4, F8, and AF4. The headset samples the collected signals at 128 Hz. The collected signals are sent to the microprocessor using Bluetooth. Data is collected for a total of 200 records where each command is recorded 50 times for an 8-second interval. However, the data collected size from all 14 channels is huge, the system has been programmed to select the 5 maximum power channels automatically using power spectral density to avoid taking a long time in processing and enhance the system performance [5][1].

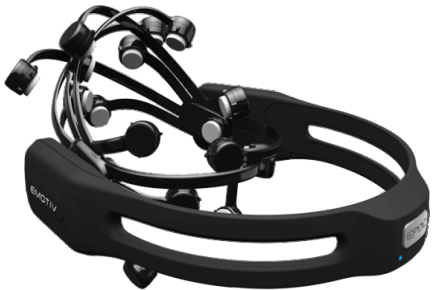

Figure 1. Emotiv Epoc Headset.

*B. Preprocessing*

The second step is to preprocess the signals to make it more suitable and noise-free for feature extraction and classification. It is important to make sure the signals acquired have minimum artifacts such as muscle movement, deep breathing or electrocardiograms (ECGs). In addition, the signals were collected in a calm environment to avoid recording noise in signals. Furthermore, the surrounding environment includes power line frequency, the headset has an inbuilt digital notch filter at 50 Hz and 60 Hz to filter power line frequency. Later, the EEG signals must be divided into different bands such as Delta, Theta…etc. for the feature extraction process [1][6].

*C. Feature Extraction*

At this stage, the signals are noise-free but are not suitable for classification. Feature extraction is considered as one of the most crucial parts of the process where it can define the success of the system. Feature extraction aims to capture the required data embedded in the EEG signals to be able to describe a specific task, it tries to represent EEG signals by extracting special features and reduce the dimension to be more suitable for classification. There are many methods that can represent EEG signals such as Fast Fourier Transform (FFT), Short Time Fourier Transform (STFT), and Wavelet Transform (WT)…etc. EEG signals are non-stationary which means it changes with time, due to this fact FFT cannot represent non-stationary signals such as EEG. As a result, STFT is introduced which is a Time-Frequency representation STFT divides the signal into a set of segments or frames where FFT processes each frame and in the end, all results are summed up in the frequency domain to provide a time-frequency representation of the signals studied. Wavelet Transform is a time-frequency domain method. It is a critical approach for feature extraction. Wavelet divides the signals into wavelets represented as a building block by contractions, shifting, compressing and stretching which is acquired from a single prototype wavelet, known as mother wavelet. The decomposed signal leads to a set of coefficients known as wavelet coefficients. Wavelet provides a better alternative over STFT since it uses an adjustable size window and can represent non-stationary signals such as EEG. For this system, Discrete Wavelet Transform (DWT) was used for feature extraction, the DWT decomposes the signals into different frequency bands by filtering the signal using Low-pass filter and High-pass filter repeatedly which is indicated by an approximation and detailed coefficients. For this system, the acquired signals have been decomposed with dB4 into 5 levels as indicated in the figure below. The approximation coefficients or the low-pass part (A1-A5) have been used to obtain the EEG frequency range which is from Delta to Gamma [7]–[9].

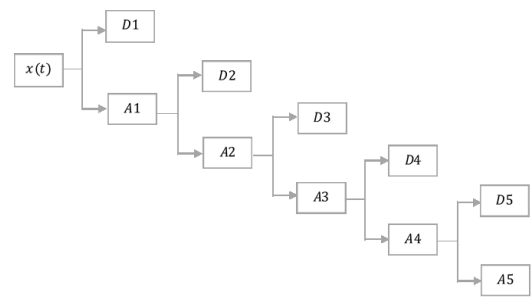

Figure 2. Five level decomposition of DWT.

*D. Classification*

The last step in EEG signal processing is to classify the signals into the desired commands such as moving to the right, left, forward, backward or stop. We have tested the accuracy of our program on 4 different classifiers such as Support vector machine (SVM), K-Nearest Neighbors (KNN), Random Forest and Artificial Neural Network (ANN). Machine learning algorithms are very powerful in regression and classification problems [10]–[13]. SVM consists of support vectors that are separated by a hyperplane which has a maximum margin between each class and the hyperplane that is being classified [14]. KNN measures the distance between the unknown and other points in the dataset and classifies the unknown to the nearest K neighbors [15]. Random forest classifier uses an ensemble learning method which aggregates the outputs from many learning algorithms to improve the classifier [16], [17]. The random forest consists of many decision trees where each decision tree votes for the class it considers that the unknown should be classified in, then based on the majority votes the unknown will be classified [18]. ANN is a method used for computation and processing of information that can mimic the behavior neurons found in the real human brain [19]. ANN has three layers which are the input layer, hidden layers, and an output layer. The layers consist of neurons that have biases and are connected using links with weights. The biases and weights are adjusted continuously until the accuracy of the classifier are very high [1]. To test our system accuracy we have assigned

80% of the data for training and 20% of data for testing for all four classifiers. Figure 3, illustrates the accuracies obtained for each classifier.

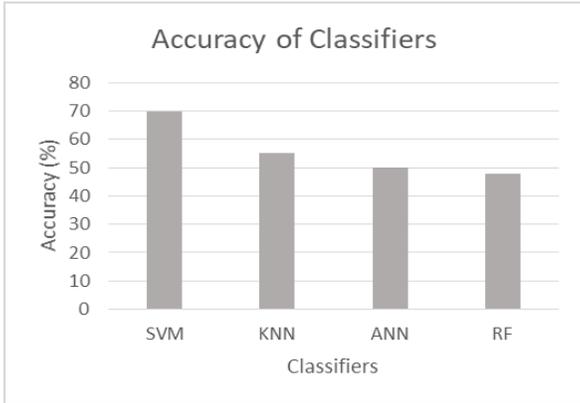

Figure 3. Accuracy of different classifiers

As shown in Figure 3, SVM is the most accurate classifier with an overall accuracy of 70% whereas KNN has an accuracy of 55%. Furthermore, ANN has an accuracy of 50% and lastly, the random forest has the lowest accuracy of 48%. Due to this fact, SVM was used as the classifier for the project.

*E. High-Level block diagram*

The block diagram in Figure 4, demonstrates how the system works. First, the data is collected using the Emotiv headset where it is sent to the microprocessor (Latte-Panda) using Bluetooth. Pre-processing, feature extraction and classification run in the microprocessor. Then based on the classified data, the command is sent to the microcontroller (Arduino Mega) to controls the prototype's movement based on the command received. We have a PC screen in the block diagram to initially run the system and visualize the processing of the system only.

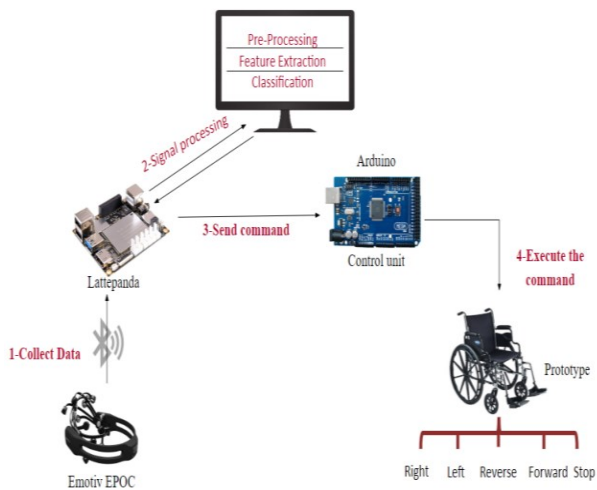

Figure 4. Block diagram.

## IV. SOFTWARE DESIGN

The project depends greatly on the software part, where the signal processing part holds the most weight. We have used different software in our project, and they are:

*a) Emotiv BCI*

This software allows the user to train using the headset, in which it displays a virtual cube for testing thought actions such as neutral, relax, pull, push, rotate...etc. to be familiar with the headset.

*b) EmotivPRO*

The Emotiv headset is capable of acquiring real-time raw EEG data of all channels and send it to a personal computer via Bluetooth at a rate of 128 Hz to display it as a waveform using EmotivPRO software. The raw data acquired can be exported to a .edf file or .csv file format to be imported for further analysis.

*c) Emotiv App*

Emotiv app manages all the interactions between the Emotiv headset, other Emotiv applications, and the PC. It is used to connect the Emotiv headset to the PC and to obtain information about the user's Emotiv account [20].

*d) PyCharm*

PyCharm is an integrated development environment (IDE) for python. It is developed by JetBrains. PyCharm is a popular IDE for python since it has many features that ease the coding process for programmers. We have written our code using Python since it has many libraries for signal processing and machine learning and free to use [21].

*e) Arduino IDE*

Arduino IDE (Integrated development environment). It is used for writing the code to control the hardware prototype [22].

## V. HARDWARE DESIGN

For the hardware, as seen in Figure 5, we demonstrated the concept of our project on a three-wheel car prototype. All the components are attached to the prototype such as the Arduino Mega, H-bridge, Latte-panda, power bank for power supply and two ultrasonic sensors to the car one at the back and another at the front for obstacle detection and to prevent crashes.

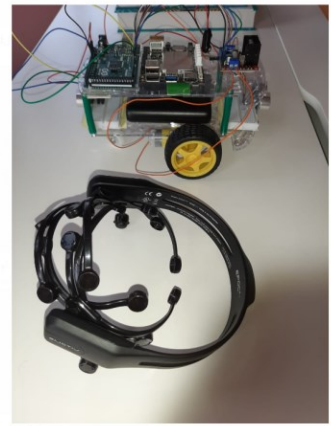

Figure 5. Prototype.

## VI. Experiment results

The user is required to sit calmly for the 8-second interval to record the required command, the data is being collected using Emotiv Epoc headset where the EEG signals then go through pre-processing, feature extraction and classification to execute the command. We have used Discrete Wavelet transform (DWT) for feature extraction and Support vector machine (SVM) for classification. We used python for writing the program software and Arduino to control the movement of the prototype depending on the command passed to it from the classifier. We have tested our system for 20 trials in real-time and obtained an overall accuracy of 68% and the accuracies for each command are shown in the Table 1. The most accurate command is Stop command while the least accurate is Reverse command.

| Commands | Number of correct trials out of 20 | Accuracy |
|---|---|---|
| Left | 14 | 70% |
| Right | 14 | 70% |
| Stop | 17 | 85% |
| Forward | 13 | 65% |
| Reverse | 10 | 50% |
| Overall Accuracy | | 68% |

Table 1. Accuracy Table of all commands

## VII. Conclusion and future work

In conclusion, the purpose of the project is to help paralyzed people to be more dependent on themselves by providing a solution to aid them in their mobility problems. As a result, this paper demonstrates an EEG based mind-controlled wheelchair as a solution to aid them. The wheelchair is controlled using thought by wearing an EEG headset called Emotiv Epoc headset that communicates using Bluetooth. The report explains the method used in implementing the project with all the software and hardware parts included. Nevertheless, the system is not ideal and definitely, there is room for improvement. The project can be further improved by adding a GPS system for navigation, converting thoughts to voice to communicate with others. As for safety improvements, the system could have a water sensing mechanism to avoid puddles and an edge detection mechanism could be added to avoid falling off cliffs. Furthermore, adjusting the speed range by thought can also be added.

## Acknowledgment

The authors thank the University of Sharjah for supporting this project.